\documentclass[useAMS,usenatbib,usegraphicx]{mn2e}

\title[Dynamical parallax of $\sigma$~Ori~AB]{Dynamical parallax of
$\sigma$~Ori~AB: mass, distance and age} 
\author[Jos\'e A. Caballero]{Jos\'e Antonio Caballero$^{1}$\thanks{Investigador
Juan de la Cierva at the~UCM.}\\ 
$^{1}$Dpto. de Astrof\'{\i}sica y Ciencias de la Atm\'osfera, Facultad de
Ciencias F\'{\i}sicas, Universidad Complutense de Madrid, \\ 
E-28040 Madrid, Spain. E-mail: caballero@astrax.fis.ucm.es}
\begin{document}

\date{Accepted 2007 October 18. Received 2007 October 01; 
in original form 2007 August 21}

\pagerange{\pageref{firstpage}--\pageref{lastpage}} \pubyear{2007}

\maketitle

\label{firstpage}

\begin{abstract}
The massive OB-type binary $\sigma$~Ori~AB is in the centre of the very
young $\sigma$~Orionis cluster.
I~have computed the most probable distances and masses of the binary for
several ages using a dynamical parallax-like method. 
It incorporates the $BVRIH$-band apparent magnitudes of both components, precise
orbital parameters, interstellar extinction and a widely used grid of stellar
models from the literature, the Kepler's third law and a $\chi^2$
minimisation.
The derived distance is 334$^{+25}_{-22}$\,pc for an age of 3$\pm$2\,Ma;
larger ages and distances are unlikely.
The masses of the primary and the secondary lie on the approximate intervals
16--20 and 10--12\,$M_\odot$, respectively.
I~also discuss the possibility of $\sigma$~Ori~AB being a triple system at
$\sim$385\,pc.
These results will help to constrain the properties of young stars and substellar
objects in the $\sigma$~Orionis~cluster.  
\end{abstract}


\section{Introduction}
\label{introduction}

The Trapezium-like system {$\sigma$~Ori}, that illuminates the encolure of
the {Horsehead Nebula}, is the fourth brightest star in the young Ori~OB~1~b
association. 
The multiple system is composed of at least five early-type stars
(Burnham 1892; Greenstein \& Wallerstein 1958; van Loon \& Oliveira 2003;
Caballero 2007b).
The two hottest components, $\sigma$~Ori~A and B (O9.5V and B0.5V), are
separated by only $\sim$0.25\,arcsec and were for a long time ``the most
massive visual binary known'' (${\mathcal M}_{\rm A} + {\mathcal M}_{\rm B}\sim
25+15~M_\odot$; Heintz 1974). 
Although the binary has not yet completed a whole revolution, the orbital
parameters are relatively well determined (Hartkopf, Mason \& McCalister 1996;
Heintz 1997; Horch et~al. 2002).
It has been suggested that $\sigma$~Ori~AB is a hyerarchical triple, being
the primary a short-period, double-line spectroscopic binary (Frost \& Adams
1904; Henroteau 1921; Miczaika 1950; Bolton 1974; Morrell \& Levato 1991). 
% with $\Delta V_r \approx$ 250\,km\,s$^{-1}$ 
However, a large amount of accurate, comprehensive spectroscopic investigations
have failed to confirm this hypothesis (Heard 1949; Conti \& Leep 1974;
Humphreys 1978; Bohannan \& Garmany 1978; Garmany, Conti \& Massey 1980;
Sim\'on-D\'{\i}az \& Lennon, priv.~comm.).  
% Henroteau F. 1921, Pub. Dominion Obs., Ottawa, 5, 45
% Heard J. F., 1949, ApJ, 109, 185

The $\sigma$~Ori system is located in the centre of the well-known
$\sigma$~Orionis open cluster. 
The proper motions, radial velocities and spacial distribution of stars in
this cluster strongly suggests a physical association between $\sigma$~Ori
itself and the young cluster (Zapatero Osorio et~al. 2002a; Caballero 2007a,
2007c -- see also Jeffries et al. [2006], who discovered a second older and
kinematically and spacially distinct population).  
Because of its youth, comparative nearness and low extinction, the
cluster has become the richest hunting ground for brown dwarfs and
planetary-mass objects in the whole sky (B\'ejar et~al. 1999; Zapatero Osorio
et~al. 2000, 2002b, 2007). 
There is plentiful material in the literature about this cluster,
covering topics like the initial mass function down to a few Jupiter masses
(B\'ejar et~al. 2001; Gonz\'alez-Garc\'{\i}a et~al. 2006; Caballero et~al.
2007), jets and Herbig-Haro objects (Reipurth et~al. 1998; Andrews et~al. 2004),
the frequency of accretors and discs (Zapatero Osorio et~al. 2002a; Oliveira,
Jeffries \& van Loon 2004;  Kenyon et~al. 2005; Oliveira et~al. 2006;
Hern\'andez et~al. 2007; Caballero et~al. 2007), the X-ray emission from young
objects (Walter et~al. 1997; Sanz-Forcada et~al. 2004; Franciosini, 
Pallavicini \& Sanz-Forcada 2006) or their photometric variability (Caballero
et~al. 2004; Scholz \& Eisl\"offel 2004).
The most used values of heliocentric distance and age of the
$\sigma$~Orionis cluster are $d \sim$ 360\,pc and $\sim$3\,Ma (Brown, de Geus
\& de Zeeuw 1994; Perryman et~al. 1997; Zapatero Osorio et~al. 2002a; Oliveira
et~al. 2002). 
There is a consensus in the literature that the cluster is younger than
8\,Ma and older than 1\,Ma.
There is, however, a strong divergence of opinion on the
heliocentric distance. 
Caballero (2007a) compiled determinations in the literature of the distance to
the $\sigma$~Orionis cluster from the 352$^{+166}_{-168}$\,pc from {\em
Hipparcos} parallax to almost 500\,pc from colour-magnitude diagrams.

Apart from the uncertainties of theoretical isochrones at very young ages,
the derivation of the initial mass function of the cluster is strongly affected
by the uncertainty in the actual age and heliocentric distance (Jeffries
et~al. 2006; Caballero et~al. 2007).
There are other investigations that require a precise age determination, such as
the evolution of the angular momentum due to discs and stellar winds
(Eisl\"offel \& Scholz 2007), disc dissipation (Hern\'andez et~al. 2007) 
and evolution of hot massive stars. % Mokiem et al. 2007 
In this Letter, I~revisit a well known method for distance determination:
the dynamical parallax (e.g. Russell 1928).
I~apply it to the $\sigma$~Ori~AB binary using state-of-the-art data and
tools to determine its mass, age and heliocentric~distance.

\section{Analysis and results}

The determination of the dynamical parallax of a binary of known
orbital period, $P$, and angular semimajor axis, $\alpha$, uses the Kepler's
third law and a power-law mass-luminosity relation (e.g. Reed 1984).
In this Letter, I~instead use the grids of theoretical models from the Geneva
group (Schaller et~al. 1992).
On the contray to other widely used grids, like those by the Lyon and Padova
groups (Baraffe et~al. 1998; Girardi et~al. 2000), the Geneva grids tabulate
absolute magnitudes in a large number of passbands and ages down to 1\,Ma and
are valid up to very high masses (i.e. ${\mathcal M} >$ 10\,$M_\odot$).
Although there are more binaries and binary candidates in the cluster
(Caballero 2005; Kenyon et~al. 2005; Caballero et~al. 2006 and references
therein), $\sigma$~Ori~AB is the only pair whose orbital parameters are known.
Here, I~have employed the parameters given by Hartkopf et~al. (1996). 
The almost face-on orbit is characterised by a long period ($P$ =
155.3$\pm$7.5\,a), a close angular separation ($\alpha$ =
0.2642$\pm$0.0052\,arcsec) and a low eccentricity ($e$ = 0.051$\pm$0.015).
The orbital parameters are consistent with those of Heintz (1997;
$P$ = 158\,a, $\alpha$ = 0.265\,arcsec, $e$ = 0.06).

Using the standard units $M_\odot$, a (annum)
%\footnote{``Annum'' is a Latin noun meaning year.  
%The International Standard ISO~31--1, the International System of Units, the
%International Astronomical Union and the National Institute of Standards and
%Technology recommend the use of the unit symbol ``a'' instead of ``yr''.
%See, e.g., {\tt http://www.iau.org/Units.234.0.html}.} 
% the "IAU Style Manual" by G.A. Wilkinson, Comm. 5, in IAU Transactions XXB (1987)
and AU for ${\mathcal M}_{\rm A} + {\mathcal M}_{\rm B}$, $P$ and the physical
semimajor axis $a$, respectively, the Kepler's third law takes the simple
expression $({\mathcal M}_{\rm A} + {\mathcal M}_{\rm B}) P^2 = a^3$.  
Accounting for $a = d \tan{\alpha} \approx d \alpha$, where $d$ is the
heliocentric distance in parsecs, and replacing the values of $P$ and $\alpha$
from Hartkopf et~al. (1996), then the third law for $\sigma$~Ori~AB can be
written as:

\begin{equation}
\label{K3law}
{\mathcal M}_{\rm A} + {\mathcal M}_{\rm B} = 7.45~10^{-7} d^3
\end{equation}

\noindent (${\mathcal M}_{\rm total} \equiv {\mathcal M}_{\rm A} + {\mathcal
M}_{\rm B}$). 
At the {\em Hipparcos} distance $d$ = 352\,pc, the total mass of the
binary would be about 32\,$M_\odot$, that is less than the classical
value of ${\mathcal M}_{\rm total}\sim$ 40\,$M_\odot$, but is consistent
with the value of
${\mathcal M}_{\rm total} \sim$ 30\,$M_\odot$ estimated by Caballero~(2007a).

\begin{table}
  \centering
  \caption{Photometry of $\sigma$~Ori~A and B from the literature.}
  \label{photometry}
  \begin{tabular}{@{}llccr@{}}
  \hline
Band	&	& $\sigma$~Ori~A& $\sigma$~Ori~B& References 			\\ % 
  \hline
$B$	& [mag]	& 3.85$\pm$0.05	& 5.18$\pm$0.05	& Ca07a		\\ % 
$V$	& [mag]	& 4.10$\pm$0.03	& 5.34$\pm$0.10	& tBr00    	\\ % 
$R$	& [mag]	& 4.15$\pm$0.04	& 5.49$\pm$0.13	& tBr00    	\\ % 
$I$	& [mag]	& 4.41$\pm$0.04	& 5.66$\pm$0.16	& tBr00    	\\ % 
$H$	& [mag]	& 4.81$\pm$0.10	& 6.02$\pm$0.10	& Ca06		\\ % 
  \hline
\end{tabular}
\end{table}

I~tabulate in Table~\ref{photometry} the $BVRIH$-band magnitudes of both
components in $\sigma$~Ori~AB, taken from the literature (ten~Brummelaar et~al.
2000 --tBr00--; Caballero 2006 --Ca06--; Caballero 2007a --Ca07a--).
Except for the $B$-band measurement, that was estimated from the Tycho-2 $B_T
V_T$ magnitudes, all the data come from adaptive optics observations.
The {\em Hipparcos} catalogue also tabulated the non-standard $H_P$-band
magnitudes.
The magnitudes and colours of both stars correspond to what it was expected of
an O9.5V and a B0.5V at $d \sim$ 350\,pc.

I~have computed through a simple minimisation method which are the most
probable heliocentric distances for several cluster ages. 
In particular, I~have looked for the minima of the following chi-square
distributions:  
%$\chi^2(d,{\mathcal M}_{\rm A})_{\rm age,Z}$

\begin{equation}
\chi^2(d,{\mathcal M}_{\rm A},{\mathcal M}_{\rm B}) =  
\chi^2_{\rm A}(d,{\mathcal M}_{\rm A}) + 
\chi^2_{\rm B}(d,{\mathcal M}_{\rm B})
\label{chi2=chi2A+chi2B}
\end{equation}

\noindent (for fixed ages and metallicities), where $\chi^2_{\rm A}$ and
$\chi^2_{\rm B}$ are:

\begin{equation}
\chi^2_{\rm A} = \sum 
\frac{(m_{\lambda,{\rm A}}-m_{\lambda,{\rm A}}^*)^2}{\delta m_{\lambda,{\rm
A}}^2},~
\chi^2_{\rm B} = \sum 
\frac{(m_{\lambda,{\rm B}}-m_{\lambda,{\rm B}}^*)^2}{\delta m_{\lambda,{\rm
B}}^2}
\end{equation}

\noindent ($\lambda \equiv B,~V,~R,~I,~H$).
$m_{\lambda,[{\rm A,B}]}$ and $\delta m_{\lambda,[{\rm A,B}]}$ are the
{\em observed} apparent magnitudes and corresponding uncertainties of
$\sigma$~Ori~A and B in Table~\ref{photometry}, and
$m^*_{\lambda,[{\rm A,B}]} = m^*_{\lambda,[{\rm A,B}]} (d,{\mathcal M}_{[{\rm
A,B}]})$ are the {\em theoretical} apparent magnitudes that a hypothetical star
of mass ${\mathcal M}_{[{\rm A,B}]}$ would have at an heliocentric distance $d$.
To compute $m^*_{\lambda,[{\rm A,B}]}$, I~have used:
($i$) the {\em theoretical} absolute magnitudes $M_{\lambda,[{\rm A,B}]}$ from
the basic grids of non-rotating stellar models with solar metallicity (Z =
0.020), overshooting and OPAL opacities of the Geneva~group, 
($ii$) the colour excess $E(B-V)$ = 0.05\,mag of $\sigma$~Ori~AB from Lee
(1968) and
($iii$) the interstellar extinction law parameters $A_\lambda / A_V$ and $R_V$
from Rieke \& Lebofski (1985).
In detail, I~have used the grids with standard mass loss $\dot{\mathcal
M}$ and ages 1.0, 2.0, 3.2, 4.9 and 10.0\,Ma (Schaller et~al. 1992 -- Sc92) and
high mass loss  $2 \times \dot{\mathcal M}$ and age 3.2\,Ma (Meynet et~al. 1994
-- Me94).
The models do not provide data for other ages less than 32\,Ma.
Caballero (2006) measured an average solar metallicity of solar-like stars in
the $\sigma$~Orionis cluster, [Fe/H] = 0.0$\pm$0.1\,dex, which justifies the use
of $Z$~=~0.020. 

%______________________________________________Fig.
\begin{figure}
\centering
\includegraphics[width=0.49\textwidth]{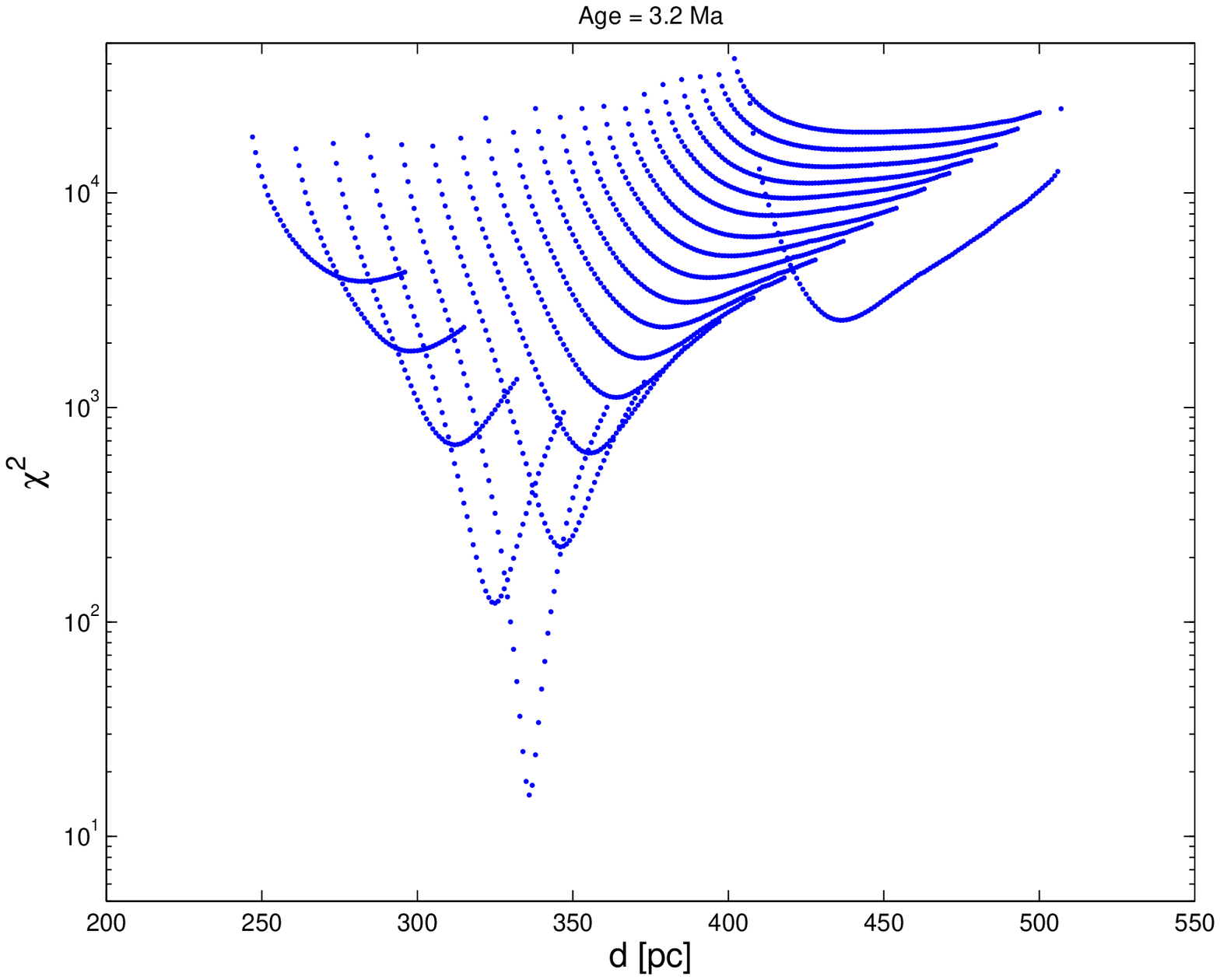}
\includegraphics[width=0.49\textwidth]{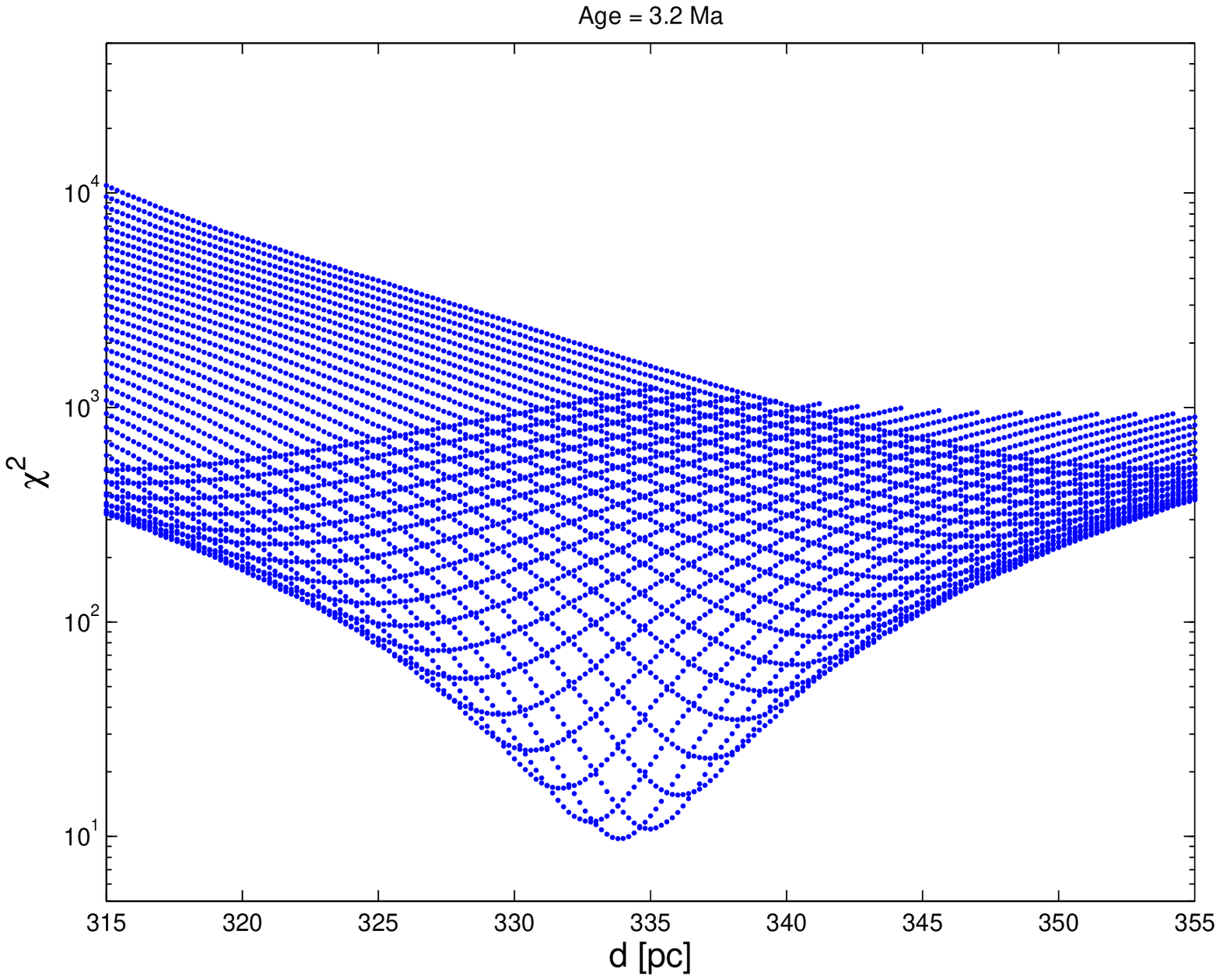}
\caption{$\chi^2$ vs. heliocentric distance for several masses of
$\sigma$~Ori~A.
The dotted lines indicate curves of constant mass.
The basic grid of models is from Schaller et~al. (1992), the age is 3.2\,Ma
and the metallicity is $Z$ = 0.020.
{\em Top window:} whole interval of heliocentric distance in steps of
1\,pc from 200 to 550\,pc and of mass of the primary in steps of 2\,$M_\odot$
from 10 to 50\,$M_\odot$. 
The minimum $\chi^2$ is for ${\mathcal M}_{\rm A}$ = 18\,$M_\odot$ ($d$ =
336\,pc).  
{\em Bottom window:} same as top window, but for intervals of heliocentric
distance in steps of 0.2\,pc and of mass of the primary in steps of
0.2\,$M_\odot$. 
The minimum in $\chi^2$ is better constrained (cf. Table~\ref{bestfits}).}
% ab01-04.m
\label{chi2d}
\end{figure}

For a fixed age (and metallicity), there is a one-to-one correspondence in the
grids between $M_{\lambda,[{\rm A,B}]}$ and ${\mathcal M}_{\rm [A,B]}$ if both A
and B stars are in the main sequence. 
For each heliocentric distance, there is only one corresponding total mass
of the hypothetical binary, ${\mathcal M}_{\rm total} \propto d^3$
(Eq.~\ref{K3law}).
If the mass of the primary, ${\mathcal M}_{\rm A}$, is fixed, then the mass of
the secondary is obtained from the simple expression ${\mathcal M}_{\rm B}
(d,{\mathcal M}_{\rm A}) = {\mathcal M}_{\rm total} (d) - {\mathcal M}_{\rm A}$.
To sum up, there is a value of $\chi^2$ for each trio $(d,{\mathcal M}_{\rm
A},{\rm age})$.
Because of the known spectral types of $\sigma$~Ori~A and~B, I~conservatively
imposed that the masses of the primary and of the secondary should lie on the
intervals 10\,$M_\odot \le {\mathcal M}_{\rm A} \le$ 50\,$M_\odot$ and
1.5\,$M_\odot \le {\mathcal M}_{\rm B} \le {\mathcal M}_{\rm A}$, respectively.
These constraints speed up the minimization but do not influence
the results (see~below).

Fig.~\ref{chi2d} illustrates the $\chi^2$ minimisation for an age of 3.2\,Ma and
standard mass loss.
Top and bottom windows display different coverage resolutions of the 
$(d,{\mathcal M}_{\rm A})$ plane.
The results for normal and high mass loss for 3.2\,Ma are almost identical.
The plots for the other ages except for 10.0\,Ma are quite similar, with the
minimum shifted along the X axis (distance). 
There are no solutions (i.e. the masses of A and B simultaneouly satisfy
Eq.~\ref{K3law} and the above constraints) for $d \la$ 250\,pc and $d 
\ga$ 500\,pc.  
In addition, the 4.9\,Ma models do not provide solutions for $d \ga$ 370\,pc. 
Finally, there are no solutions at all for 10.0\,Ma.
At this age, the primary has left the main sequence and got much brighter
than the secondary.

The values of $(d,{\mathcal M}_{\rm A})$ that minimise $\chi^2$ for a given age
and mass loss are provided in Table~\ref{bestfits}.
The uncertainties in the values of $d$ account for the error bars in the 
photometric data in Table~\ref{photometry} and in the orbital
parameters $P$ (5\,\%) and $\alpha$ (2\,\%) in Hartkopf et~al. (1996), and the
size of the steps in the high resolution minimisation ($\Delta d$ = 0.2\,pc).
The uncertainty in the masses is set to the step size, 
$\Delta {\mathcal M}_{\rm A}$ = 0.2\,$M_\odot$.
The results would be identical if no mass constraints were set.
For an age interval of 3$\pm$2\,Ma, the corresponding heliocentric distance
interval is $d$ = 334$^{+25}_{-22}$\,pc.
Distances larger than 400\,pc and less than 290\,pc are less likely for the
1--5\,Ma age range; 
distances larger than 450\,pc are highly unlikely for all ages.
%(e.g. de Zeeuw et~al. 1999).
Finally, I~have determined the most probable masses of A and B for the
{\em Hipparcos} parallax distance ($d$ = 352\,pc): ${\mathcal M}_{\rm A}$ =
21.4\,$M_\odot$, ${\mathcal M}_{\rm B}$ = 12.0\,$M_\odot$ for an age of
1\,Ma. 
The minimum $\chi^2$ for 1\,Ma is an order of magnitude smaller than for
the other ages tested, meaning that a binary age older than 1\,Ma would be
unlikely at the {\em Hipparcos} distance.

\begin{table}
  \centering
  \caption{Best fits of the dynamical parallax of $\sigma$~Ori~AB.}
  \label{bestfits}
  \begin{tabular}{@{}lccccc@{}}
  \hline
Mass		      		& Age    & $d$  	 & ${\mathcal M}_{\rm A}$& ${\mathcal M}_{\rm B}$& $\chi^2$	 \\ % 
loss		      		& [Ma]   & [pc] 	 & [$M_\odot$]   & [$M_\odot$]   &    		\\ % 
\hline
$\dot{\mathcal M}$	      	& 1.0    & 346$\pm$13	 & 20.1$\pm$0.2  & 11.7$\pm$0.2  & 8.70 	\\ % 
$\dot{\mathcal M}$	      	& 2.0    & 337$\pm$13	 & 18.3$\pm$0.2  & 11.1$\pm$0.2  & 9.38 	\\ % 
$\dot{\mathcal M}$	      	& 3.2    & 334$\pm$13	 & 17.6$\pm$0.2  & 10.8$\pm$0.2  & 9.76 	\\ % 
$\dot{\mathcal M}$	      	& 4.9    & 325$\pm$13	 & 16.0$\pm$0.2  & 10.2$\pm$0.2  & 10.56 	\\ % 
$\dot{\mathcal M}$	      	& 10.0   & ...  	 & ...  	 & ...  	 & ...  	\\ % 
$2 \times \dot{\mathcal M}$ 	& 3.2    & 333$\pm$13	 & 17.4$\pm$0.2  & 10.8$\pm$0.2  & 9.72 	\\ % 
  \hline
\end{tabular}
\end{table}

\section{Discussion}

The theoretical effective temperatures that correspond to the optimal fits lie
on the intervals $T_{\rm eff}$ = 30.4--34.6\,kK for the primary and $T_{\rm
eff}$ = 25.2--27.5\,kK for the secondary. 
The hottest temperatures are for the youngest ages. 
These values are consistent with the expected $T_{\rm eff}$ for O9.5V and
B0.5V stars, respectively (e.g. O9.5V: 30--35\,kK -- Popper 1980; Gulati,
Malagnini \& Morossi 1989; Castelli 1991; Vacca, Garmany \& Shull 1996;
Martins, Schaerer \& Hillier 2005), and with previous measurements of the
$T_{\rm eff}$ of $\sigma$~Ori~A (30.0--33.0\,kK -- Morrison 1975; Underhill
et~al. 1979; Morossi \& Crivellari 1980; Repolust et~al. 2005).
The corresponding theoretical gravities ($\log{g} \sim$ 4.00--4.22) are also
normal for class V at such temperatures. 

The minima of $\chi^2$ in Table~\ref{bestfits} are very sensitive to the
variations of heliocentric distance and of mass: 
on the one hand, at fixed mass and age, a fluctuation of $d$ of barely
30\,pc results in a change of three orders of magnitude in $\chi^2$;
on the other hand, at fixed distance and age, a fluctuation of ${\mathcal
M}_{\rm A}$ of barely 5\,$M_\odot$ results in a change of almost two orders of
magnitude in $\chi^2$.
The minima of $\chi^2$ are, however, quite unresponsive to the variations of the
age between 1.0 and 4.9\,Ma (see last column in Table~\ref{bestfits}).
A younger age gives a slightly better fit results and that favours a
slightly larger distance ($d \sim$ 350\,pc). 
The results do not strongly suggest a younger age of 1\,Ma, but they are useful
in excluding an older age. 
The absence of a solution at 10\,Ma agrees with previous upper limits on the
ages of the {Ori~OB~1~b} association from the presence of very early-type stars
in the main sequence (Blaauw 1964) and of the $\sigma$~Orionis cluster
from spectral synthesis surrounding the Li~{\sc i} $\lambda$670.78\,nm line
(Zapatero Osorio et~al.~2002a). 

The derived distance interval for 3$\pm$2\,Ma, $d$ = 334$^{+25}_{-22}$\,pc,
is consistent with the canonical distance to the $\sigma$~Orionis cluster of $d
\sim$ 360\,pc, but is difficult to conciliate with the distance of 440\,pc
for 2.5\,Ma that Sherry, Walter \& Wolk (2004) used.
The derived distance interval also deviates from very recent determinations of
the distance to some elements in the {Ori~OB~1} complex.
In particular, Terrell, Munari \& Siviero (2007), using the eclipsing
spectroscopic binary {VV~Ori} close to {Alnilam} ($\epsilon$~Ori) in Ori~OB~1~b,
and Sandstrom et~al. (2007), employing the Very Large Baseline Array in the
{Orion Nebula Cluster} in {Ori~OB~1~a}, have determined very accurate
heliocentric distances of 388--389\,pc (see also Menten et~al. 2007). 
These values are also lower than the classical distance to the
{Ori~OB~1} complex of $\sim$440\,pc from average {\em Hipparcos} parallax
(Brown, Walter \& Blaauw 1999; de Zeeuw et~al. 1999). 
Because of projection effects and the large physical size of Ori~OB~1
(Reynolds \& Ogden 1979), $\sigma$~Ori could be easily contained within the
complex.
Given their kinematic and spacial association, the $\sigma$~Ori system and the
young $\sigma$~Orionis cluster are likely at the same heliocentric distance and
also age, if one assumes that massive and low mass star formation in a cluster
is coeval (Prosser et~al. 1994; Massey, Johnson \& Degioia-Eastwood 1995;
Stauffer et~al. 1997; see, however, Sacco et~al. 2007). 
Recently, it has been suggested that the $\sigma$~Orionis cluster is
actually kinematically distinct from the {Ori~OB~1~b} association (Jeffries
et~al. 2006), just as {25~Ori} is distinct from Ori~OB~1~a (Brice\~no
et~al.~2007). 

If $\sigma$~Ori~AB were a hyerarchical triple, as described in
Section~\ref{introduction}, it would be located at a larger heliocentric
distance.
The hypothetical companion to $\sigma$~Ori~A, to which I~tentatively call
{$\sigma$~Ori~F}, would be 0.5\,mag fainter than the primary in the 370--493\,nm
interval according to Bolton (1974).
This wavelength interval corresponds to the $U,~B$ Johnson bands.
Taking into account 
$m_{\rm A} = m_{\rm A+F} + 2.5 \log{\left( 1 + 10^{-\frac{m_{\rm A}-m_{\rm F}}{2.5}} \right)}$
and the difference $B_{\rm A+F}-B_{\rm B}$ in Table~\ref{photometry}, then
the apparent magnitudes in the blue band of the three components would be
related to the combined magnitude $B_{\rm A+F}$ through: 
$B_{\rm A} \approx B_{\rm A+F} + 0.53$\,mag,
$B_{\rm B} \approx B_{\rm A+F} + 1.33$\,mag and
$B_{\rm F} \approx B_{\rm A+F} + 1.03$\,mag.
I~have looked for the distances and theoretical masses whose corresponding
apparent magnitudes match the $B_{\rm [A,B,F]}$ relations and the Kepler's third
law, $({\mathcal M}_{\rm A} + {\mathcal M}_{\rm F}) + {\mathcal M}_{\rm B} =
7.45~10^{-7} d^3$.
There are only a few solutions that simultaneously verify $T_{\rm eff,A}
\approx$ 30.0--33.0\,kK and $T_{\rm eff,B} \approx$ 26.0--30.0\,kK, as expected
from the spectral types of $\sigma$~Ori~A+F and $\sigma$~Ori~B.
In the triple scenario, the F component would have an intermediate temperature
between the A and B stars (roughly B0.0V) and would orbit very close to
$\sigma$~Ori~A. 
The valid solutions are found for the narrow distance interval 370--400\,pc and
only for ages between 1.0 and 4.9\,Ma.
Although distances less than 290\,pc and larger than 450\,pc are ruled out
again, the most probable distance to $\sigma$~Ori under the triple hypothesis,
$d \sim$ 385\,pc, agrees very well with those of VV~Ori and the Orion
Nebula Cluster.
Further high-resolution spectroscopic studies are needed to ascertain the
existence and characteristics of $\sigma$~Ori~F.

\section{Summary}

I~have determined the most probable heliocentric distances and masses of
the components in the young binary $\sigma$~Ori~AB.
The used methodology is an improvement of the dynamical parallax method using a
$\chi^2$ minimisation of observed and theoretical apparent magnitudes of
A~and~B. 
The values range in the intervals 325\,pc $\la d \la$ 346\,pc, 16\,$M_\odot \la
{\mathcal M}_{\rm A} \la$ 20\,$M_\odot$, 10\,$M_\odot \la {\mathcal M}_{\rm B}
\la$ 12\,$M_\odot$ for 1--5\,Ma.
Ages and distances larger than or equal to 10\,Ma and $\sim$450\,pc are excluded
from the minimisation. 
The theoretical effective temperatures of both components are
consistent with the observed spectral types.
Accounting for uncertainties in the orbital parameters and photometric data from
the literature, the derived distance interval for age = 3$\pm$2\,Ma is
$d$ = 334$^{+25}_{-22}$\,pc. 
The values of $\chi^2$ are minimum for the largest distances within this
interval, that translate into the youngest ages.
If there were a third component, $\sigma$~Ori~F, in a tight orbit to
$\sigma$~Ori~A, then the system could be at a larger heliocentric distance of
about~385\,pc.

The $\sigma$~Ori star system is in the centre of the young $\sigma$~Orionis
cluster. 
The knowledge of the age and heliocentric distance of the cluster is
fundamental for the study of the initial mass function down to the planetary
regime and the evolution of discs and angular momentum in very young~objects.

\section*{Acknowledgments}

I~thank the anonymous referee for improving comments and suggestions, and
D.~Montes for revising the manuscript.
Partial financial support was provided by the Universidad Complutense de Madrid
and the Spanish Ministerio Educaci\'on y Ciencia under grant 
AyA2005--02750 of the Programa Nacional de Astronom\'{\i}a y Astrof\'{\i}sica
and by the Comunidad Aut\'onoma de Madrid under PRICIT project S--0505/ESP--0237
(AstroCAM).

\bsp

\label{lastpage}

\end{document}